\documentclass[pra, twocolumn,showpacs,superscriptaddress,groupedaddress]{revtex4}  
\usepackage{graphicx}  
\usepackage{dcolumn}   
\usepackage{bm}        
\usepackage{amssymb}   
\usepackage{float}
\usepackage{mathtools}
\usepackage{mathrsfs}
\usepackage{amsmath}
\usepackage{bbm}
\usepackage{dsfont}

\newcommand{\fabsq}[1]{\left| #1 \right|^2}

\newcommand{\ket}[1]{\ensuremath{|#1\rangle}}
\newcommand{\bra}[1]{\langle#1|}
\newcommand{\braket}[2]{\langle#1|#2\rangle}
\newcommand{\ketbra}[2]{|#1\rangle\langle#1|}
\newcommand{\cH}{\begin{cal}H\end{cal}}

\newcommand{\bN}{\mathbb{N}}
\newcommand{\bid}{\mathds{1}}

\newcommand{\sgn}{\mathop{\textrm{sgn}}}
\newcommand{\temp}{\tau}
\newcommand{\Mop}{\hat{\mathscr{M}}}
\newcommand{\Fid}{{\cal{F}}}
\newcommand{\kB}{k_\textrm{B}}
\newcommand{\lin}{\textrm{lin}}

\newcommand{\fref}[1]{Fig.~\ref{#1}}
\newcommand{\sref}[1]{Sec.~\ref{#1}}

\newcommand{\eref}[1]{Eq.~(\ref{#1})}
\newcommand{\Fref}[1]{Figure~\ref{#1}}

\usepackage{ulem} 
\usepackage{color}

\begin{document}

\title{Quantum Coherent Control via Pauli Blocking}

\author{Tom Dowdall}
\address{Department of Physics, University College Cork, Cork, Ireland}

\author{Albert Benseny}
\address{Quantum Systems Unit, Okinawa Institute of Science and Technology Graduate University, 904-0495 Okinawa, Japan}

\author{Thomas Busch}
\address{Quantum Systems Unit, Okinawa Institute of Science and Technology Graduate University, 904-0495 Okinawa, Japan}

\author{Andreas Ruschhaupt}
\address{Department of Physics, University College Cork, Cork, Ireland}

\begin{abstract}
Coherent quantum control over many-particle quantum systems requires high fidelity dynamics. One way of achieving this is to use adiabatic schemes where the system follows an instantaneous eigenstate of the Hamiltonian over timescales that do not allow transitions to other states. This, however, makes control dynamics very slow. Here we introduce another concept that takes advantage of preventing unwanted transitions in fermionic systems by using Pauli blocking: excitations from a protected ground state to higher-lying states are avoided by adding a layer of  {\it buffer} fermions, such that the {\it protected} fermions cannot make a transition to higher lying excited states because these are already occupied. This allows to speed-up adiabatic evolutions of the system. We do a thorough investigation of the technique, and demonstrate its power by applying it to high fidelity transport, trap expansion and splitting in ultracold atoms systems in anharmonic traps.
Close analysis of these processes also leads to insights into the structure of the orthogonality catastrophe phenomenon.
\end{abstract}

\maketitle

\section{Introduction}
\label{sec:intro}

Preparation of  and coherent control over many-particle quantum states requires quantum engineering techniques that lead to high fidelities.
Adiabatic processes, where the system follows an eigenstate of the time-dependent Hamiltonian, are known to allow for this; however
they require that the Hamiltonian is varied sufficiently slowly in order to avoid transitions to other eigenstates~\cite{messiah_book}.
This leads to long process times and leaves the system vulnerable to decoherence, reducing also the possible repetition rates of the process.

How quickly or slowly an eigenstate can be followed depends roughly on the distance to the next closest-lying eigenstate~\cite{messiah_book}.
Therefore, one strategy for expediting adiabatic processes is to adjust the instantaneous speed of the process with respect to the size of the instantaneous level gap such that the transition probability to unwanted eigenstates remains small during the whole process~\cite{Guerin_2002, Guerin_2011, Garaot_2015}.
This, however, requires the knowledge of the energy eigenspectrum during the whole process. 

In recent years, a number of techniques to speed up adiabatic processes have been developed under the name ``shortcuts to adiabaticity''~\cite{sta_review_1, sta_review_2}. One example of these techniques relies on the implementation of an additional counter-diabatic Hamiltonian, which is designed to compensate for any excitations that appear during the finite time  evolution process, such that the system does not leave the eigenstates of the original Hamiltonian~\cite{demirplak_2003,berry_2009, chen_2010b}. However, this additional Hamiltonian can be  very complicated and thus be demanding to implement experimentally.
Other shortcut techniques are based on Lewis--Riesenfeld invariant inverse engineering~\cite{chen_2010a}, which allow for a fast transfer of all initial eigenstates simultaneously to all final eigenstates (up to a phase).

Generalizing these techniques to many particle systems is not a straightforward task, as the number of degrees of freedom increases exponentially with larger particle numbers. The effects of this are well known and can be seen immediately when considering one of the most simple systems possible, namely
an ideal, spin-polarized, one-dimensional Fermi gas at low temperatures: even in the presence of almost perfect single-particle process fidelities, the overlap between two many-particle wavefunctions scales with $N^{-\alpha}$, where $\alpha$ depends on the specific nature of the change between the initial and final Hamiltonian~\cite{orthocat-1}. This is the  so-called orthogonality catastrophe (OC)~\cite{orthocat-2,orthocat-3}, which has recently been examined for systems of ultracold fermions~\cite{Goold:11,Campbell:14}

\begin{figure}[tb]
\begin{center}
\includegraphics[width = 0.47 \columnwidth]{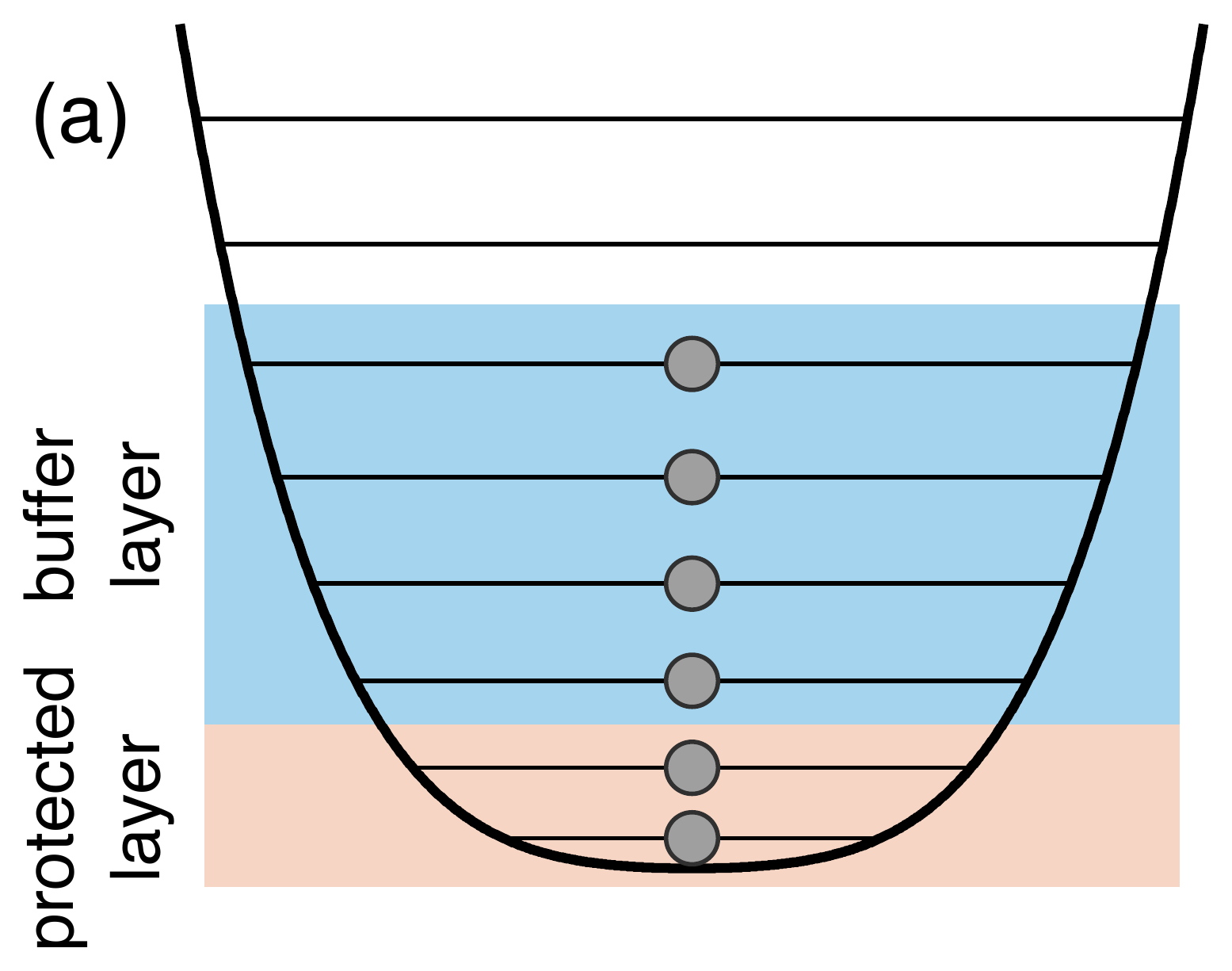}
\includegraphics[height = 0.36 \columnwidth]{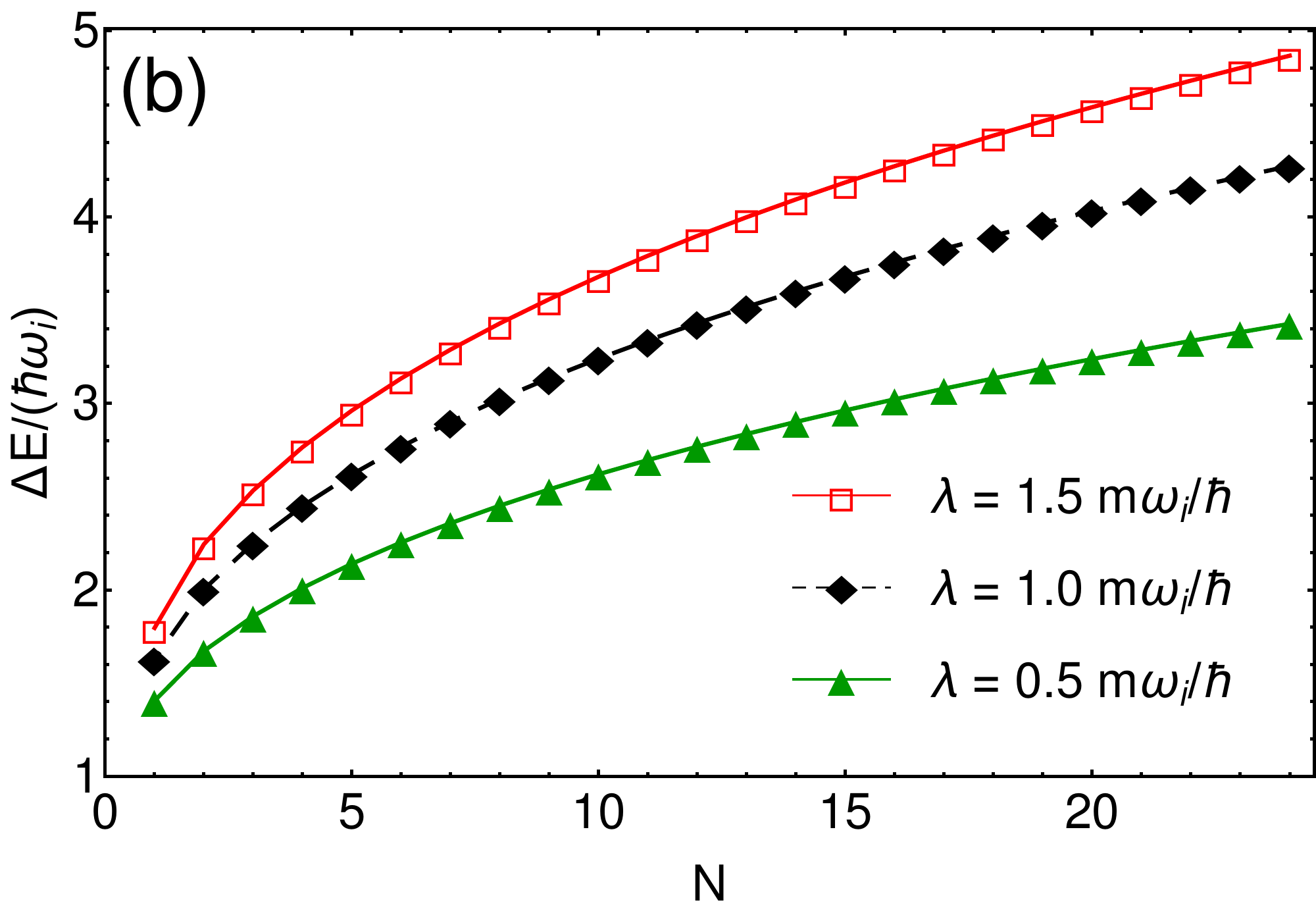}
\end{center}
\caption{
\label{Fig1:Schematic}
(a) Schematic of the key idea:
In order for the particles in the protected zone to remain in the lower energy eigenstates during a time-dependent change of the external control parameters, a buffer zone is added. The Pauli principle then prevents the protected atoms from accessing any level in the buffer zone and to access an unoccupied level above the Fermi edge requires a large amount of energy.
(b) Fermi gap $\Delta E= E_{N+1}-E_N$ versus total particle number $N$ for the anharmonic trap $V(x) = m\omega_i^2 (x^2 + \lambda x^4)/2$ for different anharmonicities $\lambda$.
}
\end{figure}

Here, however, we show that this behavior does not necessarily limit the engineering of many-particle states, as the OC does not affect all states inside a Fermi sea in the same way.
In fact, one can always find a kernel of particles that is essentially unperturbed, and whose size scales with the overall number of particles.
This is due to the fact that transitions inside the Fermi sea are forbidden by the Pauli exclusion principle and lead to the so-called Pauli blocking, which has recently been examined to engineer cold atomic systems~\cite{Busch:98, Ferrari:99,Demarco:01,OSullivan:09,Omran:2015}.

In this work we will consider a system of trapped, ultracold, spin-polarized fermionic atoms, and explore the idea of using Pauli blocking for speeding up adiabatic evolutions.
In addition to the ground state layer of particles that should be protected from making transitions we are also adding a buffer layer of particles, see Fig.~\ref{Fig1:Schematic}(a).
The basic idea now is that only the fermions close to the Fermi edge can make transitions, whereas all atoms inside the Fermi sea need significantly more energy to get excited.
Since we are only interested in the protected particles, this will allow to carry out {\it adiabatic} processes much faster, as long as the energies introduced by the dynamics do not allow for particles in the protected layer to make transitions.
Once the evolution is finished, the {\it buffer fermions} can be discarded by, for example, lowering the trap walls~\cite{Serwane_2011} or inducing spin-flips as in similar techniques for the evaporative cooling of bosons~\cite{Ketterle_1996}.
As this technique can most easily protect ground states, it is particularly well suited to prepare initial states in potentials where direct ground-state cooling is challenging.

The idea we present relies on the specific form of the energy spectrum around the Fermi edge. If the Fermi edge is close to the continuum states in a finite height potential, it is not guaranteed that the process we investigate will work. However, if the spectrum becomes increasingly sparse beyond the Fermi edge
(for example in anharmonic trapping potentials, see \fref{Fig1:Schematic}(b)),
significant speedups can be obtained. 
In fact,  in this limit the idea of Hilbert space engineering through quantum statistics is largely independent of the potential shape, i.e.~the exact form of the Hamiltonian.

Since the technique we discuss below will protect the lower motional energy states, and since the protection is done by the presence of a Fermi sea, it requires fermionic samples that are deep within the quantum degenerate regime.  For neutral atoms these can be produced routinely in laboratories worldwide these days~\cite{DeMarco:99,Partridge:06,Shin:08} and since the removal of the higher energy particles from a trap can also be done using standard techniques, we will concentrate in this paper on the control process itself.

In the following we will first introduce the system we investigate and define and discuss the process fidelity as our figure of merit. 
We will then apply the method in detail to three specific control tasks in \sref{sec:tasks}, and conclude in \sref{sec:conclusions}.


\section{System and fidelity}
\label{sec:fidelity}

\subsection{Fermion state}

We consider a gas of spin-polarized fermions that formally consists of $N_p$ particles whose state we want to protect and $N_b$ particles that form a buffer layer (see Fig.~\ref{Fig1:Schematic}(a)), so that the overall number of particles is $N=N_p+N_b$. Since at ultracold temperatures the dominant scattering interaction is of symmetric s-wave form, such gases can be efficiently described as non-interacting and they therefore form a perfect Fermi sea at zero temperature~\cite{Giorgini:08}. This also means that the time evolution of the many-particle wave function, $\ket{\Psi(t)}$, can be obtained by solving the single-particle Schr\"odinger equations for each state within the Fermi sea
\begin{equation}
	i \hbar \frac{\partial}{\partial t} \ket{\psi_i(t)} = \left[-\frac{\hbar^2}{2m}\frac{\partial^2}{\partial x^2} + V(x,t)\right] \ket{\psi_i(t)},
	\label{eq:1p_tdse}
\end{equation}
where the shape and time-dependence of the potential, $V(x,t)$, depends on the particular task that is to be implemented. The many particle wavefunction then follows from calculating the Slater determinant as
\begin{equation}
	\label{eq:multi_wf}
	\ket{\Psi(t)} = \dfrac{1}{\sqrt{N}} \sum_{\sigma \in \Pi[N]} \sgn(\sigma)\prod_{i = 1}^{N} \ket{\psi_{\sigma(i)}(t)}_{i} ,
\end{equation}
where $\Pi[N]$ consists of all the permutations of the set $\{1,\ldots,N\}$.

\subsection{Process fidelity}

In the following we will consider processes where the $N_p$ lowest eigenstates of an initial Hamiltonian are occupied by $N_p$ relevant particles and  we aim at having this subset of the Fermi sea to be undisturbed during the evolution towards the final Hamiltonian.
In order to quantify how well the process works we calculate the overlap between the evolved state at the final time $T$, $\ket{\Psi} \equiv \ket{\Psi(T)}$, and the lowest lying eigenstates $\ket{\phi_{i}} (i=1,\ldots,N_p)$ of the Hamiltonian at the end of the process.
In detail, we define the fidelity of the process as 
\begin{equation}
	\Fid = \braket{\Psi}{\Mop\Psi},
\label{fid1}
\end{equation}
where $\ket{\Psi}$ is an element of the fermionic subspace ${\cH}_F^N$ of the $N$-particle Hilbert space ${\cH}^N$
and the measurement operator $\Mop$ is defined as
\begin{align}
\label{eq:M_op}
\Mop &= \dfrac{1}{N_b!} \sum_{\tau \in \Pi[N]} \hat{M}^{(\tau(1))} \, \otimes \; \ldots \; \otimes \, \hat{M}^{(\tau(N))} ,
\\
\hat{M}^{(i)} &=
\begin{cases}
\ket{\phi_{i}}\bra{\phi_{i}} & \textrm{if} \quad i = 1,\ldots,N_p, \\
\bid  & \textrm{if} \quad i = N_p+1,\ldots,N.
\end{cases}
\end{align}
The operator $\hat{M}^{(i)}$ checks the occupation probability of the $i$-th eigenstate of the Hamiltonian, provided that $i \leq N_p$, and,
as we are not interested in the population of levels above $N_p$, $\hat{M}^{(i)}$ acts as the identity for $i > N_p$.

Let $\hat P_F$ be the projector on the fermionic subspace ${\cH}_F^N$.
For $\ket{\Psi} \in {\cH}_F^N$, we have $\Fid = \braket{\Psi}{\Mop \Psi} = \braket{\Psi}{\Mop_F \Psi}$
where $\Mop_F := \hat P_F  \Mop \hat P_F$. One can show by using the fermionic number basis states $\ket{\Phi_{\vec n}}$ that
\begin{equation}
\Mop_F = \sum_{\vec n} \ketbra{\Phi_{\vec n}}{\Phi_{\vec n}},
\end{equation}
where the sum is over all vectors $\vec n$ fulfilling $n_j = 1$ for $j=1,\ldots,N_p$ and $\sum_{j=N_p+1}^\infty n_j = N_b$.
From its structure it is clear that the operator $\Mop_F$ is a projector. This proves that always $0 \le \Fid \le 1$ as it should be for a meaningful fidelity definition.

The fidelity \eqref{fid1} can then be rewritten as (see Appendix for details)
\begin{equation}
\Fid = \sum_{U} \bigg| \sum_{\sigma \in \Pi[N_p]} \sgn(\sigma) \prod_{i=1}^{N_p} \braket{\psi_{U(\sigma(i))}(T)}{\phi_{i}} \bigg|^{2},
\label{eq:fid}
\end{equation}
where the first sum $U$ is over all
mappings $U:\{1,\ldots,N_p\} \to \{1,\ldots,N\}$ with $U(i)<U(i+1)$ for $i=1,\ldots,N_p-1$
(which can be also viewed as all subsets of cardinality $N_p$ of the set $\{1,\ldots,N\}$).
As mentioned above, the states $\ket{\psi_j(T)}$ can be obtained from the single-particle Schr\"odinger equation (\ref{eq:1p_tdse}).

From Eq. (\ref{eq:fid}), it also follows that $\Fid^{(N_b+1)} \ge \Fid^{(N_b)}$, i.e.~that $\Fid$ increases monotonically with the number of buffer particles $N_b$.
This can be seen because
\begin{align}
\Delta \Fid &= \Fid^{(N_b+1)} - \Fid^{(N_b)}\label{monotonic}\\
&= \sum_{U\setminus \tilde U} \bigg| \sum_{\sigma \in \Pi[N_p]} \sgn(\sigma) \prod_{i=1}^{N_p} \braket{\psi_{U(\sigma(i))}(T)}{\phi_{i}} \bigg|^{2} \geq 0 \nonumber
\end{align}
where $U$ are all a subset of cardinality $N_p$ of the set $\{1,\ldots,N+1\}$ and
$\tilde U$ are all a subset of cardinality $N_p$ of the set $\{1,\ldots,N\}$.
Note that from this property and the fact that $\Fid$ is bounded by $1$, we know
that the limit $\lim_{N_b\to\infty} \Fid^{(N_b)}$
must exist, but it is not necessarily 1.

\subsection{Adiabaticity and shortcuts}

Let us first look at schemes which work perfectly in the adiabatic limit, i.e., for $T \to \infty$.
In this limit one gets $\ket{\psi_j (T)} + e^{i \zeta_j (t)} \ket{\phi_j (T)}$
where the $\zeta_j$ are phases. It  immediately follows from Eq. (\ref{eq:fid}) that $\Fid = 1$.
To be more general, if $T$ is large but finite, we get that $\ket{\psi_i (T)} = e^{i \zeta_i (t)} \ket{\phi_i (T)} + \frac{1}{T} \ket{\chi_i^{(1)} (T)} + \frac{1}{T^2} \ket{\chi_i^{(2)}(T)} + \ldots$
where the phase of $\ket{\phi_i (t)}$ can be chosen in such a way that $\braket{\phi_i (T)}{\chi_i^{(1)}(T)} = 0$.
Based on this, we can make a series expansion of the fidelity in the small parameter $1/T$ as
\begin{equation}
\Fid \simeq 1 + \frac{1}{T^2} \left[ 
\alpha^{(0)}
+ \sum_{\mu=1}^{N_p} \sum_{\lambda=1}^{N_b} \fabsq{\braket{\chi_{N_p+\lambda}^{(1)}(T)}{\phi_\mu (T)}} 
\right],
\end{equation}
where $\alpha^{(0)}$ is an expression independent of $N_b$. However, it can be seen that all terms which depend on $N_b$ are always positive
and therefore improve the fidelity. This coincides with the general monotonicity of the fidelity in $N_b$ shown above.

Another special case are settings where shortcuts to adiabaticity techniques can be applied exactly,
like for example the expansion of a harmonic trap \cite{chen_2010a} or the transport in a harmonic trap \cite{torrontegui_2011}.
One can see from the above equation that one would obtain $\Fid = 1$ exactly for arbitrary numbers of particles on arbitrary timescales.
In the following, we will therefore concentrate on settings where a shortcut to adiabaticity cannot be found easily, in particular anharmonic settings.

\subsection{Temperature effects}

To extend this approach to the case of a finite temperature $\temp$, the
initial state is of canonical form and
 the probability for a specific occupation $m$ at initial time is given by
\begin{equation}
p_{m} = \frac{1}{Z} \exp\left[- \frac{1}{\kB \temp} \sum_{j=1}^{N} (E_{m (j)} - E_{j}) \right] .
\end{equation}
Here
$Z = \sum_{m} \exp\left[- \sum_{j=1}^{N} (E_{m(j)} - E_{j})/\kB \temp\right]$ is the partition function
and $\kB$ is the Boltzmann constant. The sum is over all functions $m:\{1,...N\} \to \bN$ with $m(i) < m(i+1)$, i.e.
$(m (1),\ldots,m (N))$ are the numbers of the energy eigenstates occupied by the $N$ fermions and the
$E_j$ are the ordered eigenenergies of the Hamiltonian at the initial time.
The finite-temperature fidelity will then be the average over the fidelities of the different possible permutations of the particles
\begin{equation}
\Fid = \sum_{m} p_{m} \Fid_{m},
\end{equation}
where $\Fid_{m}$ is the fidelity defined similar to the one above with just the states in $(m (1),\ldots,m (N))$
initially occupied instead of $(1,\ldots,N)$:
\begin{equation}
\Fid_{m} = \sum_{U} \bigg| \sum_{\sigma \in \Pi[N_p]} \sgn(\sigma) \prod_{i=1}^{N_p} \braket{\psi_{m(U(\sigma(i)))}(T)}{\phi_{i}} \bigg|^{2},
\end{equation}
Note that while this sum is in principle infinite, we will truncate it for its numerical evaluation at a maximal energy level chosen such that the result is practically independent from the exact level of truncation.

\begin{figure*}
\begin{center}
(a) \hspace{0.55\columnwidth} (b) \hspace{0.55\columnwidth} (c) \\
\includegraphics[height = 0.2 \textwidth]{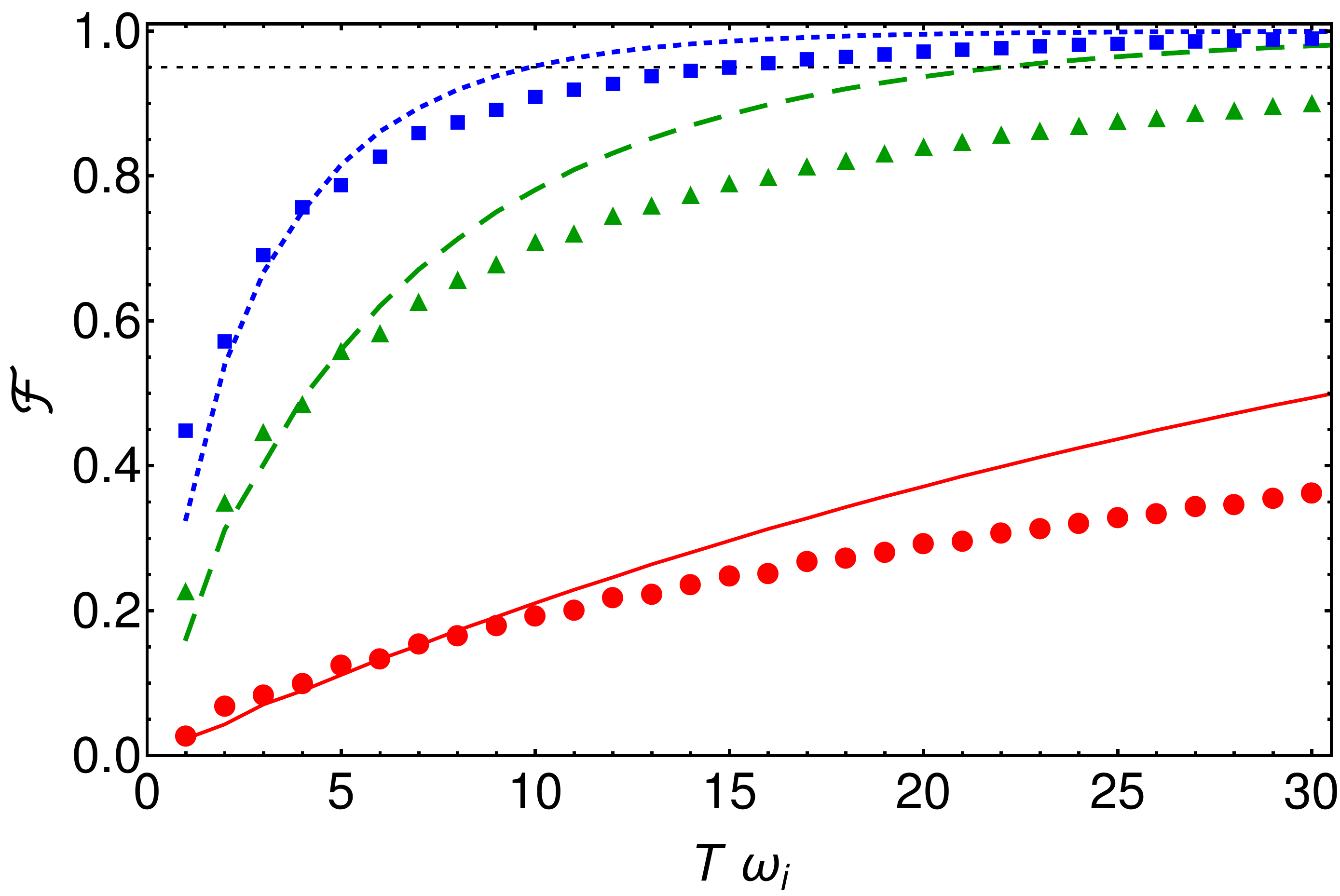}
\includegraphics[height = 0.2 \textwidth]{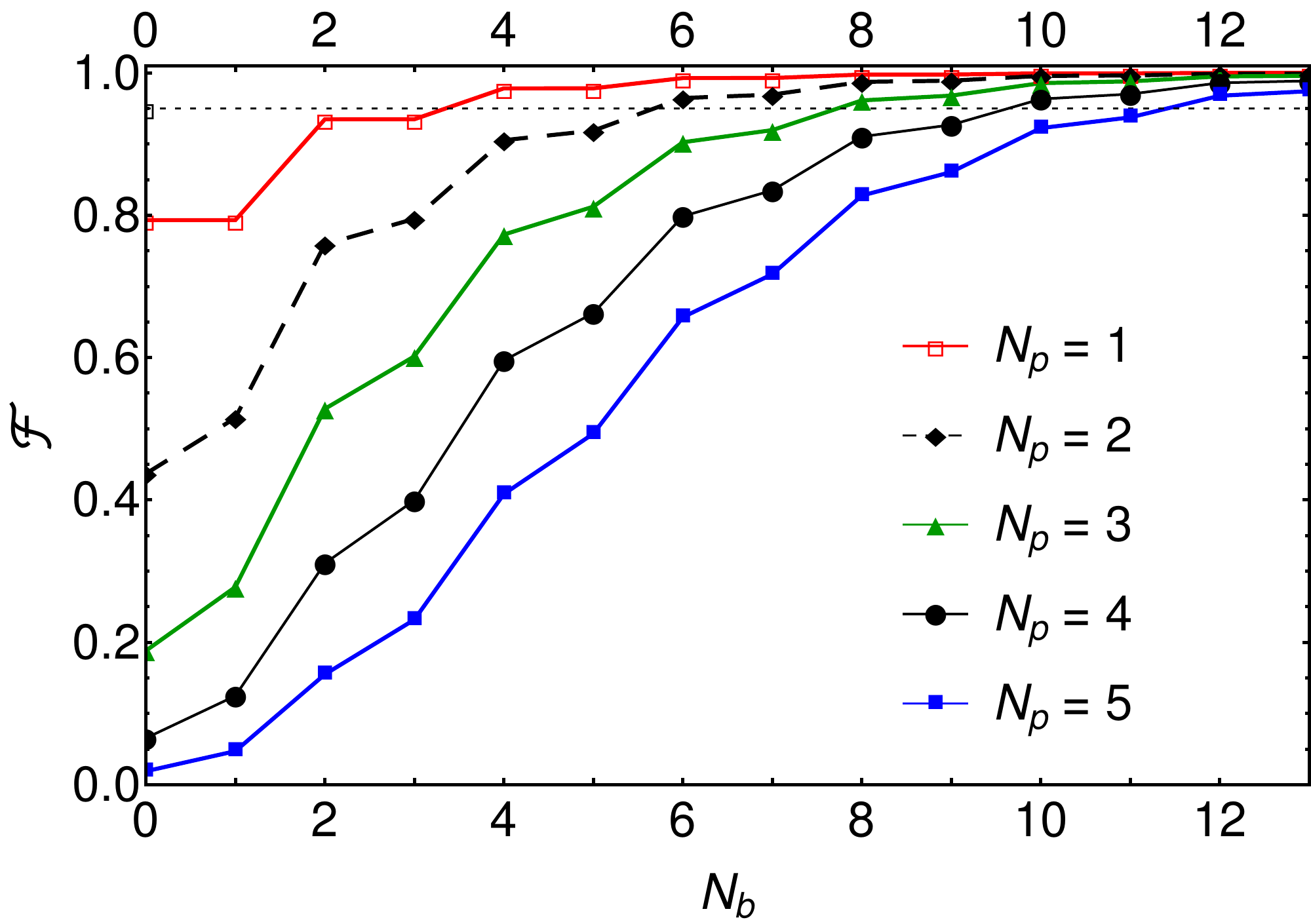}
\includegraphics[height = 0.2 \textwidth]{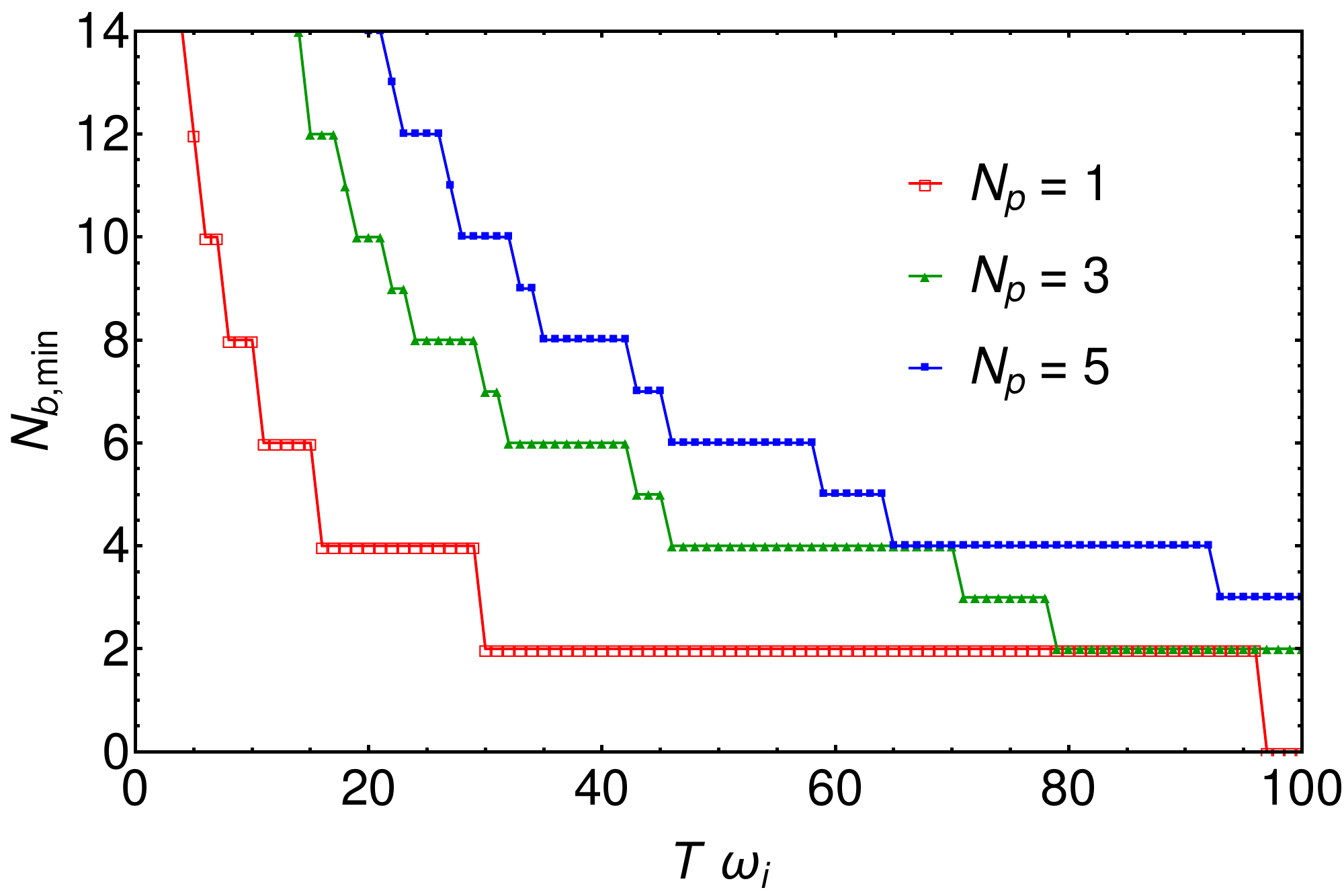}
\end{center}
\caption{
Trap expansion with $\omega_f/\omega_i = 0.01$ at a temperature $\tau = 0$.
(a) $\Fid$ versus $T$ for $N_p=2$;  lines indicate the sinusoidal scheme and the markers indicate the linear scheme.
$N_b = 0$ (red solid line/circles),
$N_b = 6$ (green dashed line/triangles),
$N_b = 12$ (blue dotted line/squares).
The horizontal black dotted line in (a) and (b) indicates $\Fid=0.95$.
(b) $\Fid$ versus $N_b$ for $T=25/\omega_i$ with the sinusoidal scheme for different $N_p$.
(c) Minimal number of buffer particles required to achieve $\Fid \geq 0.95$ versus $T$ for different $N_p$.
}
\label{Fig2:Exp_1}
\end{figure*}

\begin{figure}
\begin{center}
(a) \\
\includegraphics[height = 0.2 \textwidth]{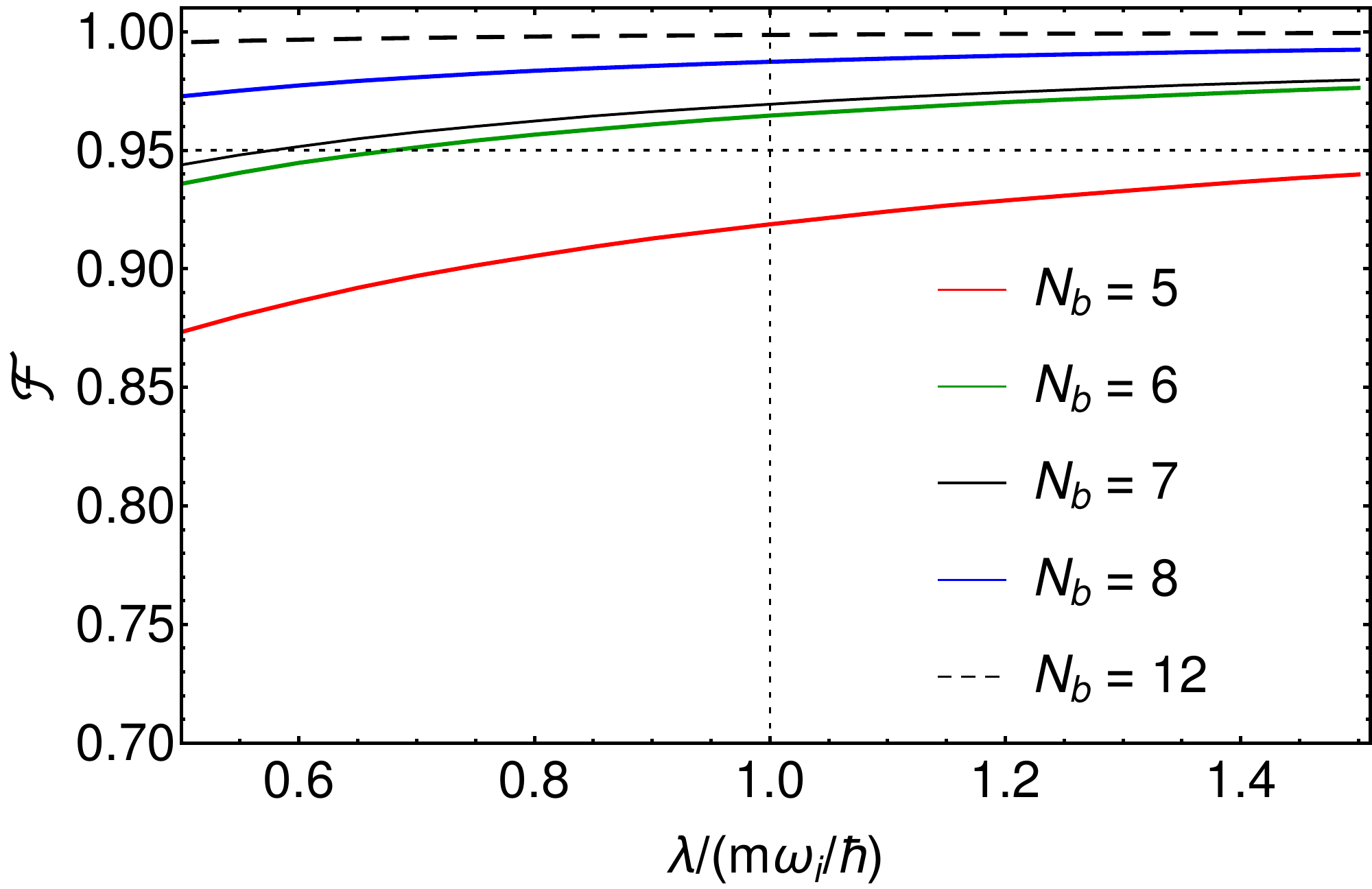} \\
(b) \\
\includegraphics[height = 0.2 \textwidth]{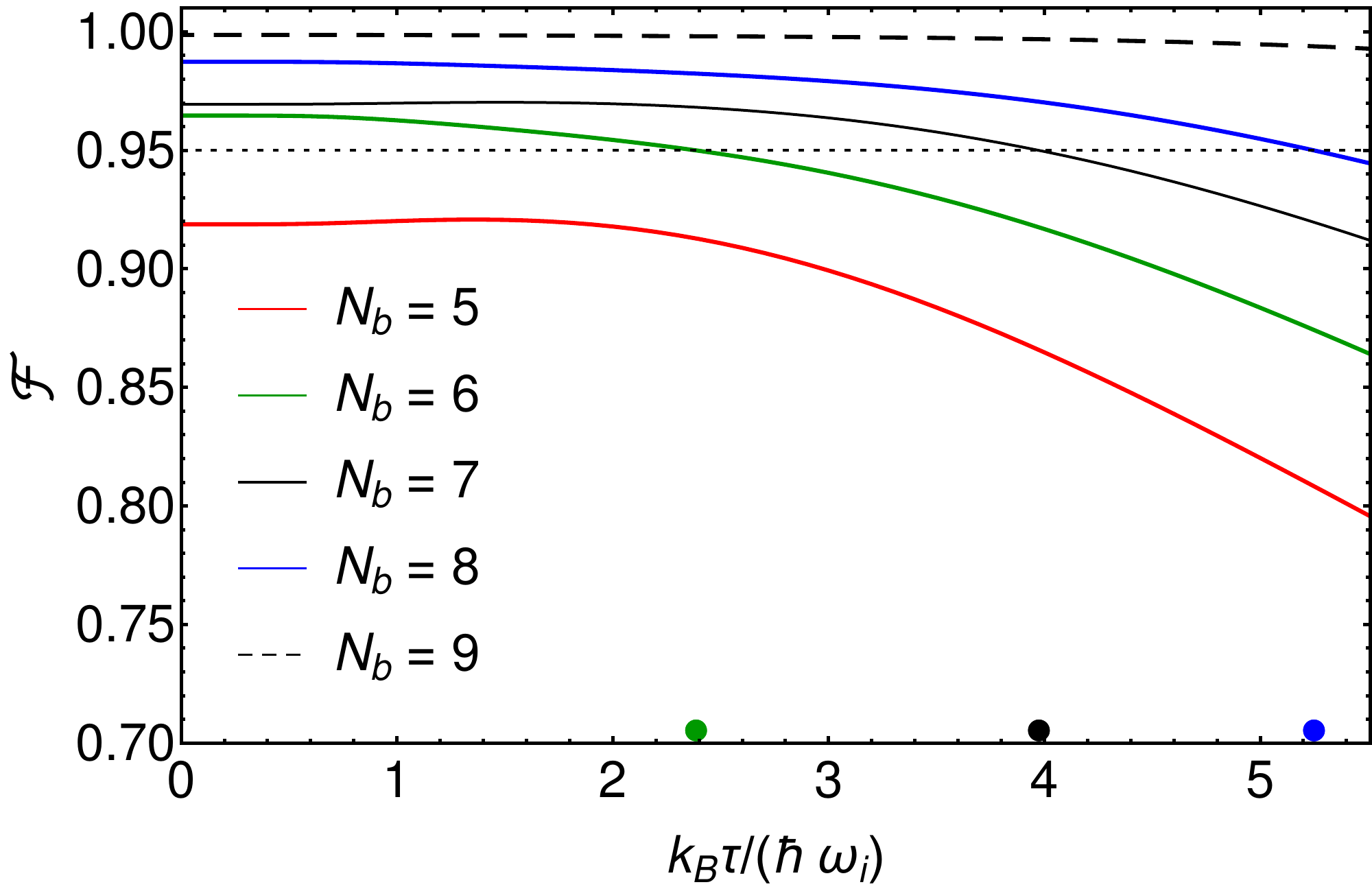}
\end{center}
\caption{
Trap expansion with a sinusoidal scheme for different number of buffer particles $N_b$.
(a) Fidelity $\Fid$ versus anharmonicity $\lambda$ at temperature $\tau=0$; the vertical line indicates $\lambda = m \omega_i/\hbar$, to allow easy comparison to \fref{Fig2:Exp_1}.
(b) Fidelity $\Fid$ versus temperature $\tau$, $\lambda = m \omega_i/\hbar$.
In both figures: 
$\omega_f/\omega_i = 0.01$, $N_p=2$ with $T = 25/\omega_i$; the horizontal line indicates a fidelity of $\Fid=0.95$;
in (b) the dots on the horizontal axis indicate when the corresponding line crosses this threshold fidelity.
}
\label{Fig3:Exp_2}
\end{figure}

\section{Control tasks}
\label{sec:tasks}

In this section we focus on particles trapped in potentials with significant anharmonicities, such that these cannot be treated as perturbations, and discuss three manipulation examples: expansion, transport, and splitting of the trap. 
For small (or zero) trap anharmonicity shortcuts for expansion and transport have been
derived \cite{sta_review_1,torrontegui_2011,torrontegui_2012a,torrontegui_2012b,STA_anharm_trans} and shortcuts related to the splitting can be found, for example, in \cite{torrontegui_2013, martinez_2013}.

This broad variety of tasks will show that, in contrast to other shortcut-to-adiabaticity protocols, 
the idea presented here is insensitive to the details of how the trap parameters are varied in time and does not require any specific time-dependence parameter functions which might be very complex and hard to implement experimentally.
The only parameter is the number of buffer particles, $N_b$, and we will show below how the fidelity depends on the size of the buffer for each of the three processes.

\subsection{Trap expansion}

We first consider the expansion of the trapping potential, which we choose to be of the form
\begin{equation}
	V(x,t) = \dfrac{m}{2} \omega(t)^2 \left( x^{2} + \lambda x^{4}\right),
\end{equation}
and in which the anharmonicity is quantified by the parameter $\lambda$.
We set $\lambda= m\omega_i/\hbar$ such that the anharmonicity is significant and far from being
just a small perturbation.

For the control task the trapping frequency $\omega(t)$ is changed from $\omega_i$ at $t=0$ to $\omega_f$ at $t=T$ and
we consider two different forms of the time-dependence, linear and sinusoidal, respectively given by 
\begin{align}
	\omega_{\lin}(t) &= \omega_{i} + (\omega_{f} - \omega_{i}) \frac{t}{T} , \\
	\omega_{\sin}(t) &= \omega_{i} + (\omega_{f} - \omega_{i}) \sin^{2}\left(\frac{\pi t}{2 T}\right).
\end{align}
The resulting fidelities $\Fid$ for both schemes are shown in \fref{Fig2:Exp_1}(a) for $N_p = 2$.
One can clearly see that adding just a small number of buffer particles leads to significantly larger $\Fid$, even at total times for which the fidelity without buffer particles was very low.
We also see that this is independent of the control scheme, underlining the fact that our method does not depend on the precise time-dependence of the control parameters.
Nonetheless, it can be seen that the sinusoidal scheme generally results in larger $\Fid$ than the linear scheme for fixed $T$ and $N_b$.
Since both schemes yield roughly similar results we will in the following focus on the sinusoidal scheme only.

The dependence of the fidelity on the number of buffer particles for different numbers of protected particles $N_p$ is shown in 
\fref{Fig2:Exp_1}(b) 
for a fixed process time of $T=25/\omega_i$.
The fidelity increases monotonically with increasing $N_b$ (for fixed $N_p$),
agreeing with the general property of the fidelity derived in \sref{sec:fidelity}.

In addition, it is interesting to note that adding an even number of particles is more effective than adding an odd number.
This can be understood by first considering the extreme case of $N_p = 1$ (red line in \fref{Fig2:Exp_1}(b)), where it can be seen that, if one add a single buffer particle to an even number of buffer particles, the process fidelity does not change. The reason for this is  that expansion is a symmetric operation with respect to the center of the trap, i.e.~the Hamiltonian commutes with the parity operator. Therefore states of different parity do not couple and for $N_p=1$ the subspace of buffer particles in odd eigenstates completely decouples from the subspace of the single, protected particle (as the ground state is even) and also from the buffer particles in even eigenstates. The fidelity then depends only on the {\it even} subspace and  adding an additional {\it odd} buffer particle
has no effect. For larger $N_p$, both subspaces are involved in the fidelity, making the situation more complex and the effect less prominent.

\fref{Fig2:Exp_1}(b) also illustrates the effect of the OC, as one can see that 
fidelities decrease dramatically with larger system sizes (larger $N_p$). However, it is also worth pointing out that in our situation this is slightly surprising, as 
due to the trap anharmonicity, the Fermi gap is bigger for larger $N_p$, see \fref{Fig1:Schematic}(b), and one could therefore expect the OC to be suppressed for larger systems at fixed $T$. 
Nevertheless, \fref{Fig2:Exp_1}(b)  clearly shows that
adding more particles to the system increases the fidelity of the relevant, lower lying many-body state, and therefore allows to {\it beat} the OC.

In all cases a fidelities $\Fid \ge 0.95$  can be achieved by adding a large enough number of buffer particles and
\fref{Fig2:Exp_1}(c) shows the relation between the process time $T$ and the minimal number of buffer particles $N_{b,\min}$
needed for achieving $\Fid \geq 0.95$ for all process times larger than $T$. It can clearly be seen that smaller $T$
must be combined with a larger number of  buffer particles, $N_b$, to result in the desired threshold fidelity.

Next, we study the effects of the potential shape and the temperature on our scheme and
start by considering the dependence on $\Fid$ for different (relevant, non-perturbative) anharmonicities $\lambda$.  
The results shown in \fref{Fig3:Exp_2}(a) confirm that this method does not require a detailed knowledge of the trapping potential, as for $N_b \ge 8$ the fidelity stays always above the threshold fidelity of $0.95$ for the whole range of $\lambda$ values shown.
In fact, we note that the fidelity increases with $\lambda$ as our scheme takes advantage of the increased energy gap at the Fermi energy for larger $\lambda$ (see again \fref{Fig1:Schematic}(b)).

Finite temperature results are shown in \fref{Fig3:Exp_2}(b)
for different numbers of buffer particles $N_b$ (with fixed $T=25/\omega_i, N_p=2, \lambda = m\omega_i/\hbar$).
and it can be seen that the scheme is quite stable under temperature perturbations.
Increasing temperatures can be compensated by increasing the number of buffer particles to achieve the same target fidelity:
$N_b$ should be increased by one to compensate for an increase in temperature of the order of $\hbar\omega_i/k_B$ (see the dots in \fref{Fig3:Exp_2}(b)).
This is also what one would expect heuristically as the ``width'' of the edge in the Fermi--Dirac distribution is of the order of $k_B \tau$
and the energy gap is of the order $\hbar\omega_i$. 
 As one might expect, the increase of the fidelity is again monotonic with increasing $N_b$ with finite temperature for the shown parameter range.

\begin{figure*}
\begin{center}
(a) \hspace{0.55\columnwidth} (b) \hspace{0.55\columnwidth} (c) \\
\includegraphics[height = 0.2 \textwidth]{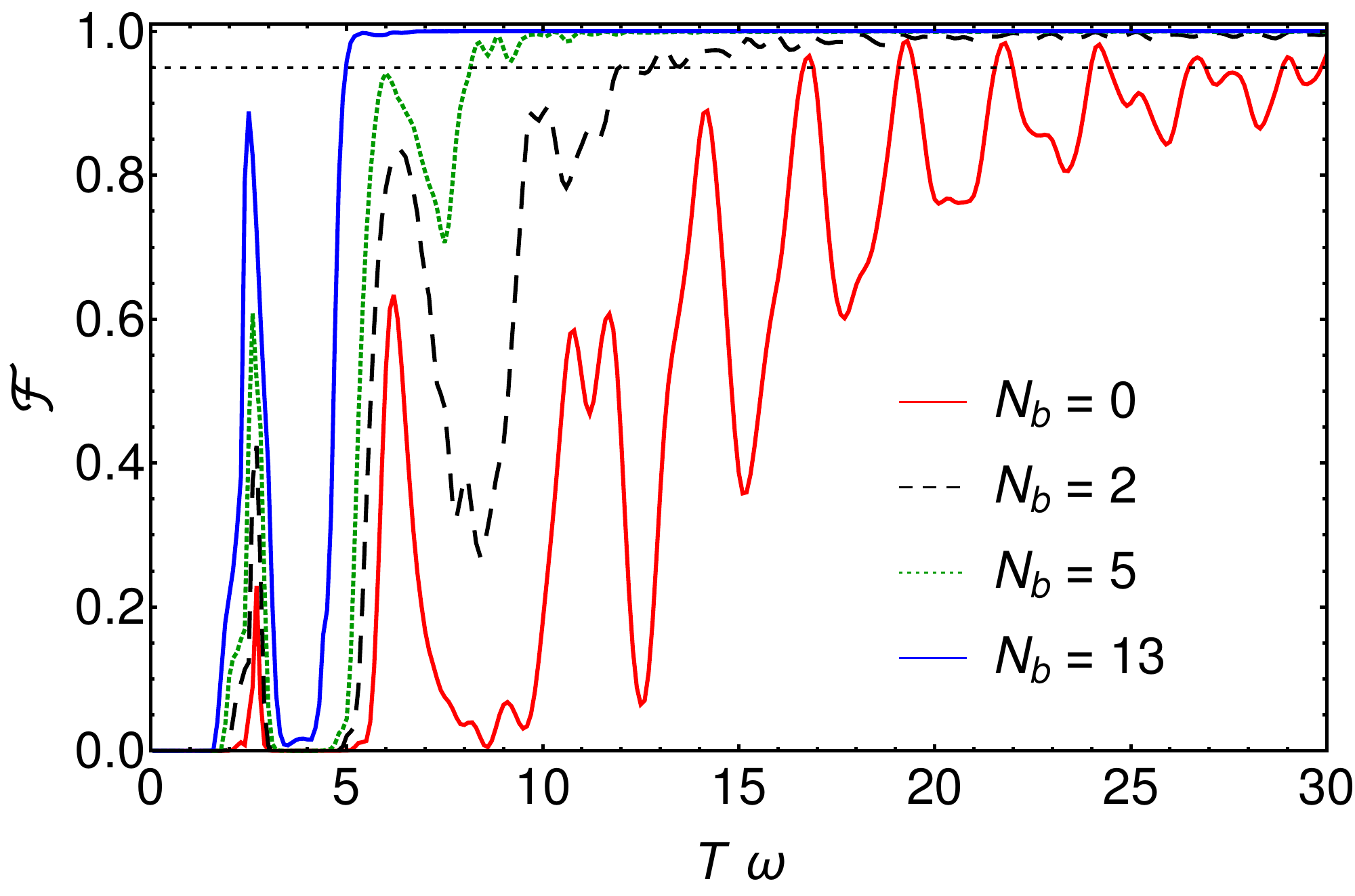}
\includegraphics[height = 0.2 \textwidth]{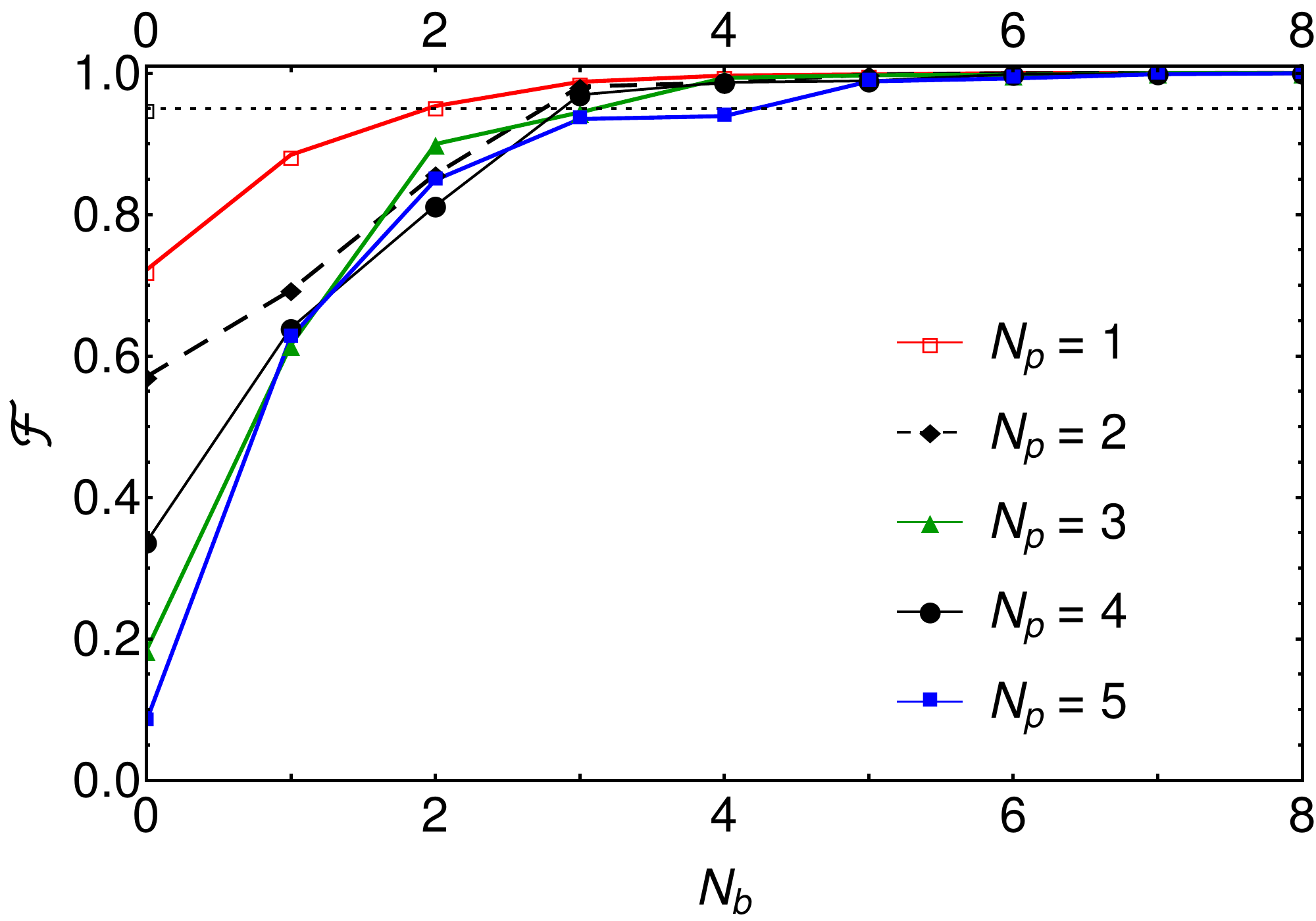}
\includegraphics[height = 0.2 \textwidth]{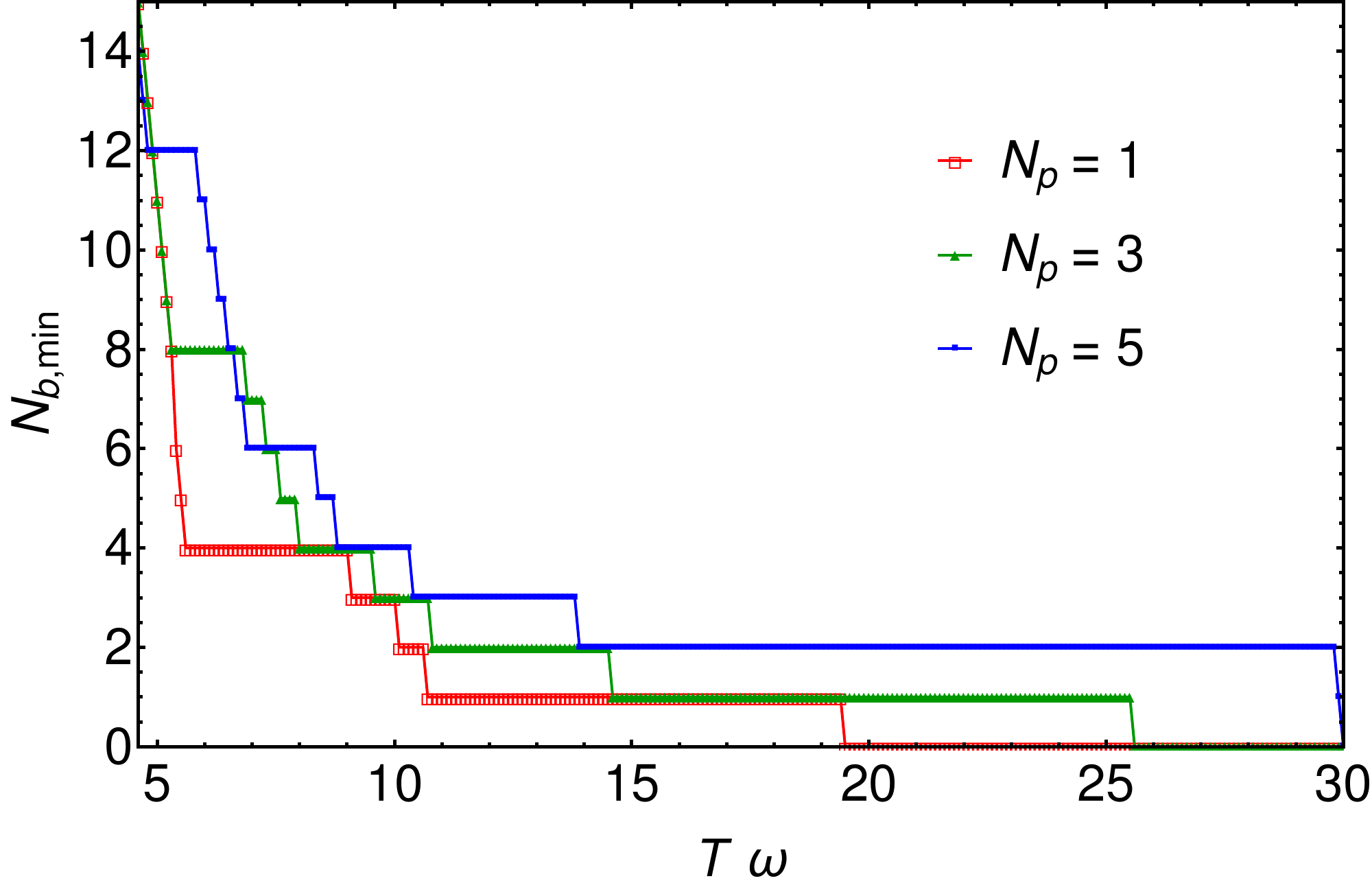}
\end{center}
\caption{
Trap transport with the sinusoidal scheme from $x_{0i} = 0$ to $x_{0f} = 90 d$ at temperature $\tau = 0$.
(a) Fidelity $\Fid$ versus process time $T$ for different $N_b$ with $N_p=2$.
(b) Fidelity versus $N_b$ for different $N_p$, $T=11.5/\omega$.
(c) Minimal buffer particles $N_{b,\min}$ versus process time $T$ for different $N_p$.
The horizontal black dotted lines in (a) and (b) indicate a fidelity of $\Fid=0.95$.
}
\label{Fig4:Trans_1}
\end{figure*}

\begin{figure}
\begin{center}
(a) \\
\includegraphics[height = 0.2 \textwidth]{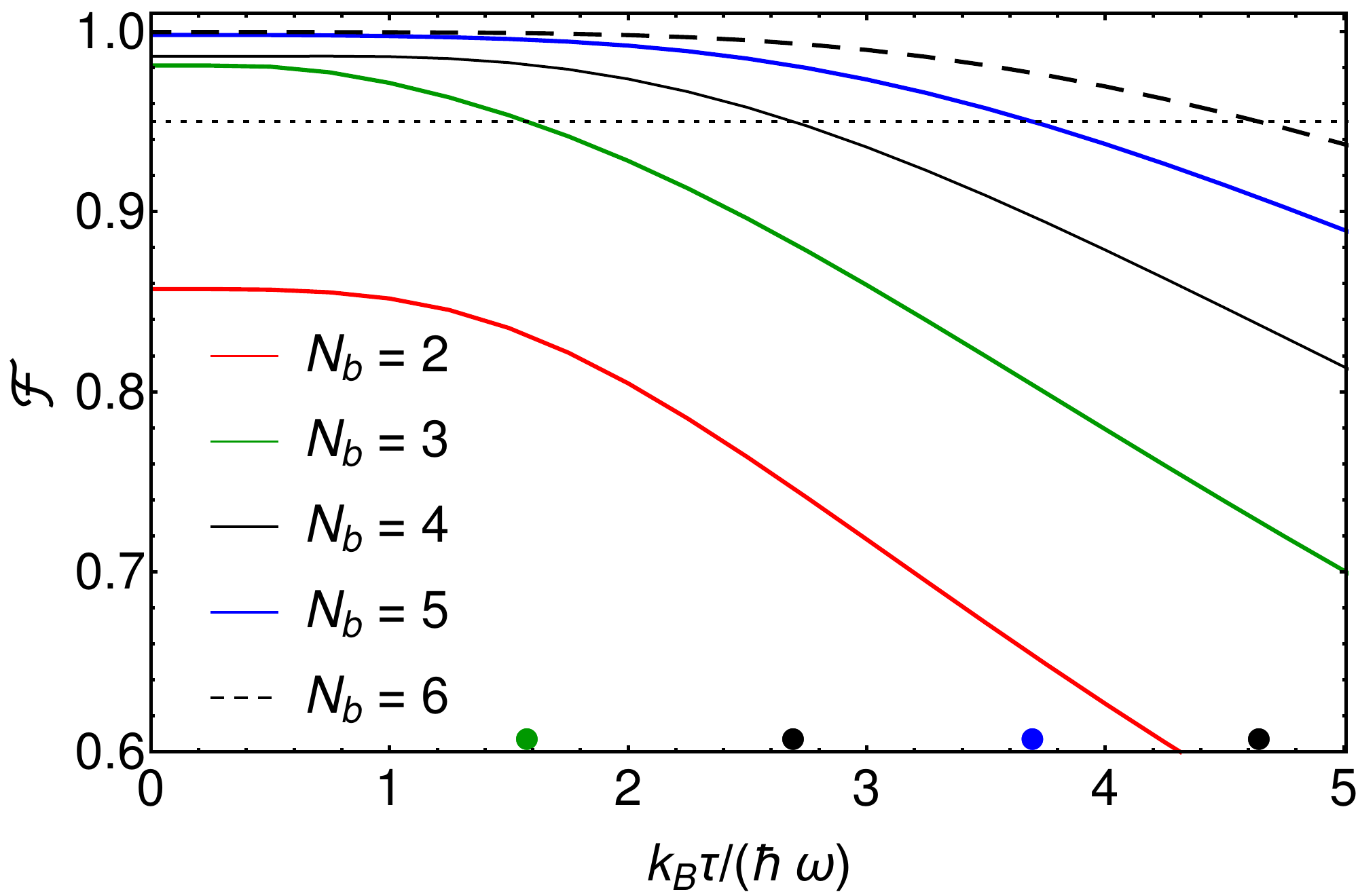} \\
(b) \\
\includegraphics[height = 0.2 \textwidth]{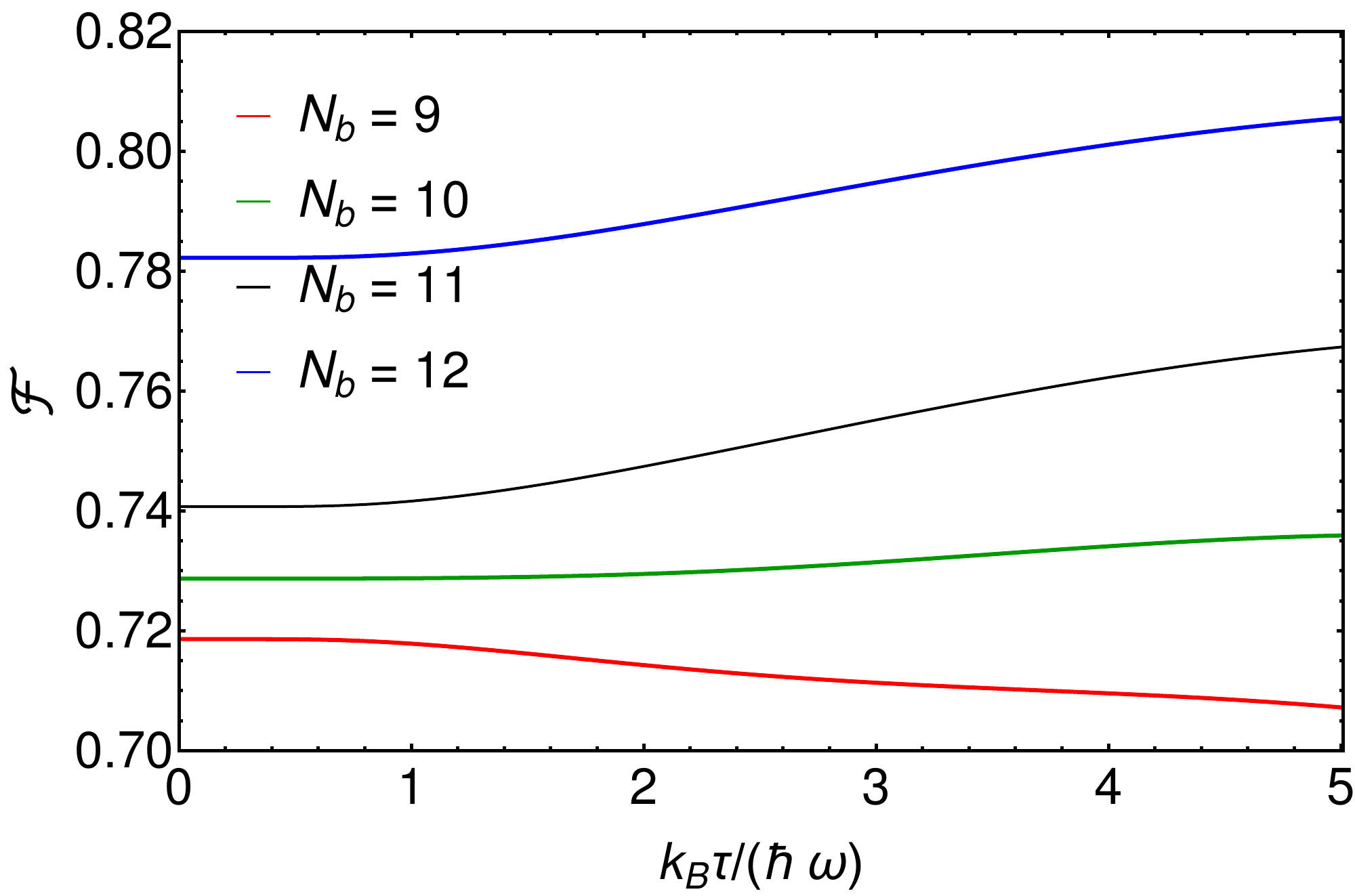}
\end{center}
\caption{
Trap transport from $x_{0i} = 0$ to $x_{0f} = 90 d$,
Fidelity $\Fid$ versus temperature $\tau$ for different $N_b$, $T \omega = 11.5$, $N_p=2$:
(a) sinusoidal scheme;
the horizontal line indicates $\Fid=0.95$, and the dots on the horizontal axis indicate when the corresponding line crosses this threshold fidelity.
(b) Linear scheme $x_0 (t) = x_{0f} t/T$.
}
\label{Fig5:Trans_2}
\end{figure}

\subsection{Transport}

The second dynamical scheme we examine is the spatial translation of the trapping potential described by
\begin{equation}
V(x,t) = \frac{1}{2} m \omega^{2} \left((x - x_{0}(t))^{2} + \lambda(x - x_{0}(t))^{4}\right) ,
\end{equation}
and we choose the movement of the trap center $x_0(t)$  between $x_i = x_0 (0)$ and $x_f = x_0 (T)$ to be of the form
\begin{equation}
x_{0}(t) = x_{i} + (x_{f} - x_{i})\sin^{2}\left(\frac{\pi t}{2 T}\right).
\end{equation}
Let $d=\sqrt{{\hbar}/{m\omega}}$, and we set $\lambda= 1/d^2$. 
The resulting fidelities $\Fid$ are shown in \fref{Fig4:Trans_1}(a) for $N_p = 2$ and one can see that,
similarly to the expansion scheme, fidelities of $\Fid \ge 0.95$ can be achieved by increasing the number of buffer particles $N_b$ instead of increasing the total time $T$.
In this case, however, the fidelities exhibit oscillations for shorter $T$, giving high fidelities for some specific final times
(similar to  \cite{couvert_2008}).

In \fref{Fig4:Trans_1}(b) we examine how the fidelity depends on the number of buffer particles for different numbers of protected particles $N_p$ (for a fixed process time $T=11.5/\omega$).
As expected, adding buffer particles $N_b$ always increases the fidelity (see again also \sref{sec:fidelity}). 
However, it is worth pointing out certain differences compared to the expansion scheme (see \fref{Fig2:Exp_1}(a)).
First, adding a single buffer particle always has  a significant effect and second, the fidelity is now not monotonic in $N_p$ (for fixed $N_b$ and $T$,  compare to \fref{Fig2:Exp_1}(b)):
all fidelity lines for the different $N_p$ cross the threshold line of $\Fid = 0.95$ given enough $N_{b}$.

\Fref{Fig4:Trans_1}(c) shows the relation between the process time $T$ and the minimal number of buffer particles $N_{b,\min}$ required to reach $\Fid \geq 0.95$ for all process times larger than or equal to $T$.
Similar to the expansion scheme, $N_{b,\min}$ goes to 0 for large enough $T$ and the required buffer is increasing for shorter
process times $T$. In addition,  $N_{b,\min}$ does not have a strong dependence on $N_p$ in the transport case.

The relation between $\Fid$ and temperature $\tau$, for different values of $N_b$ (with fixed $T=11.5/\omega$, $N_p=2$), is shown in \fref{Fig5:Trans_2}(a).
One can see that the scheme is again stable against temperature perturbation, however, for increasing temperature the number of buffer particles $N_b$ has to be increased to still achieve a fidelity $\Fid \ge 0.95$.
Again, from the dots on the horizontal axis it can be seen that $N_b$ has to be increased by one if the temperature increases by an order of
$\hbar\omega_i/k_B$.
Again, we note that for the temperatures shown there is still the monotonic increase of the fidelity with increasing $N_b$.

It is also interesting to note that the fidelity in general does not always decrease monotonically with increasing temperature.
This can be seen in \fref{Fig5:Trans_2}(b), where a linear transport scheme is considered and an increase in fidelity is visible in certain temperature ranges. The reason for this lies in the internal structure of the Fermi sea at finite temperatures, which allows for highly complex dynamics.

\begin{figure*}
\begin{center}
(a) \hspace{0.55\columnwidth} (b) \hspace{0.55\columnwidth} (c) \\
\includegraphics[height = 0.2 \textwidth]{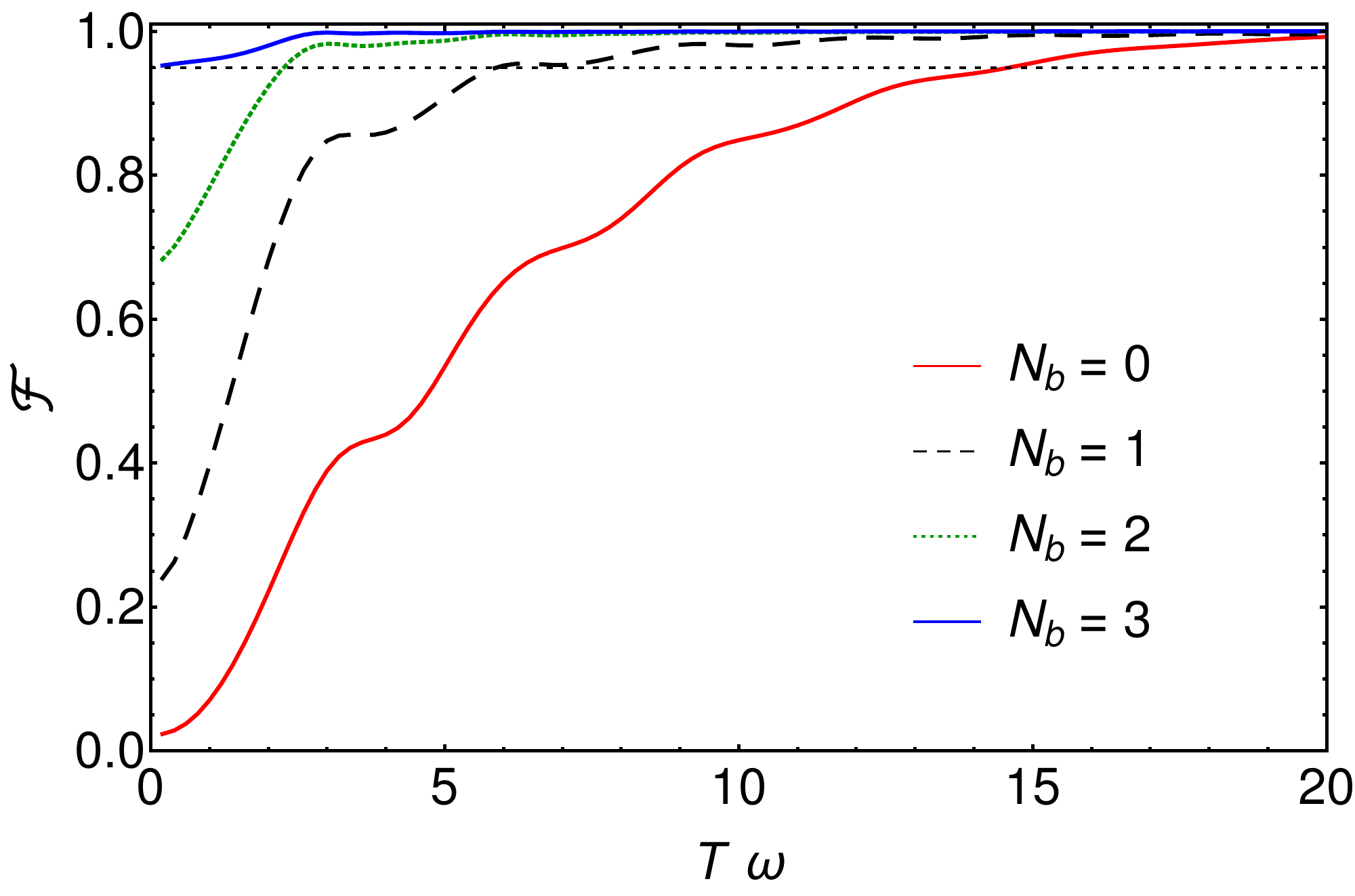}
\includegraphics[height = 0.2 \textwidth]{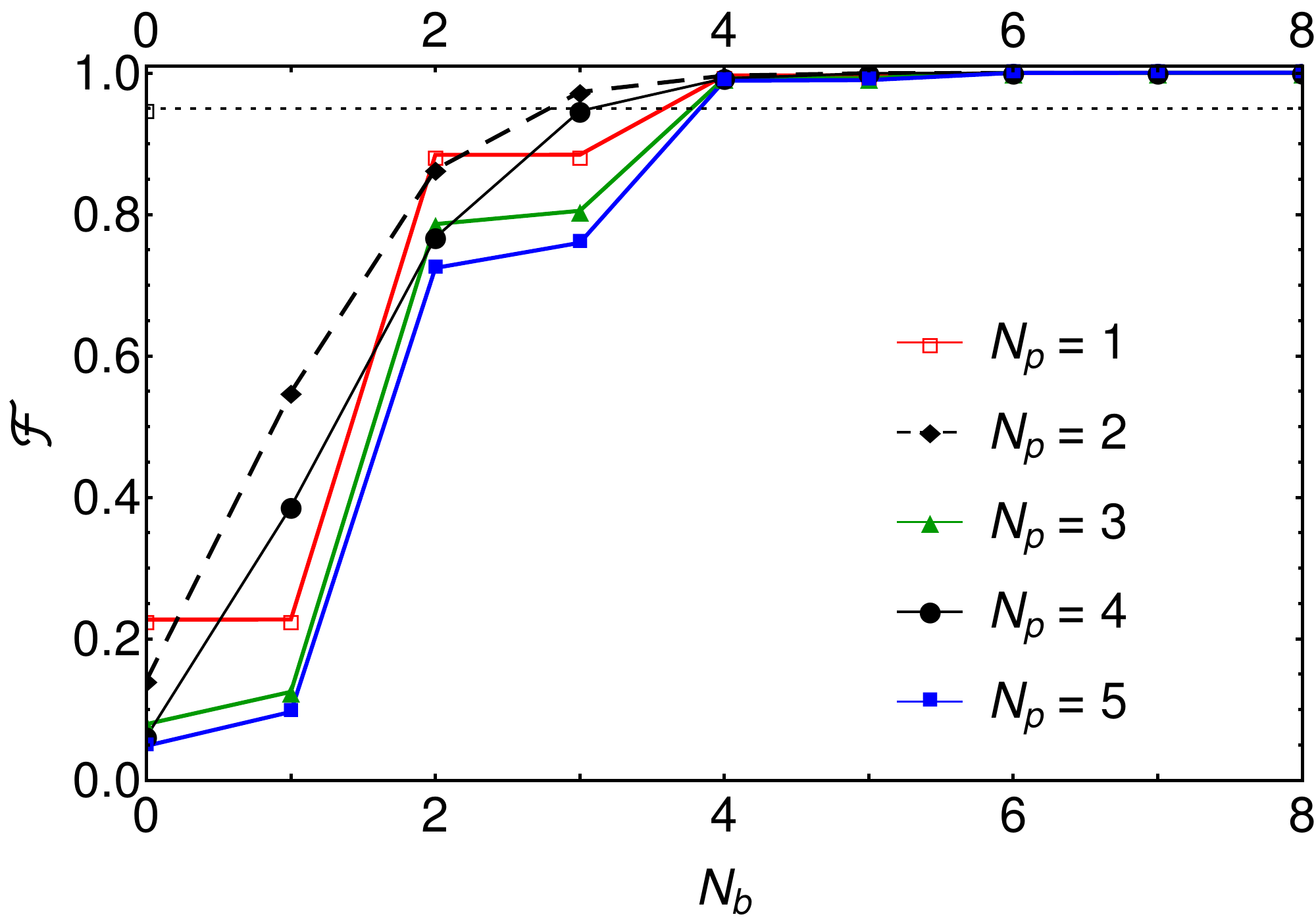}
\includegraphics[height = 0.2 \textwidth]{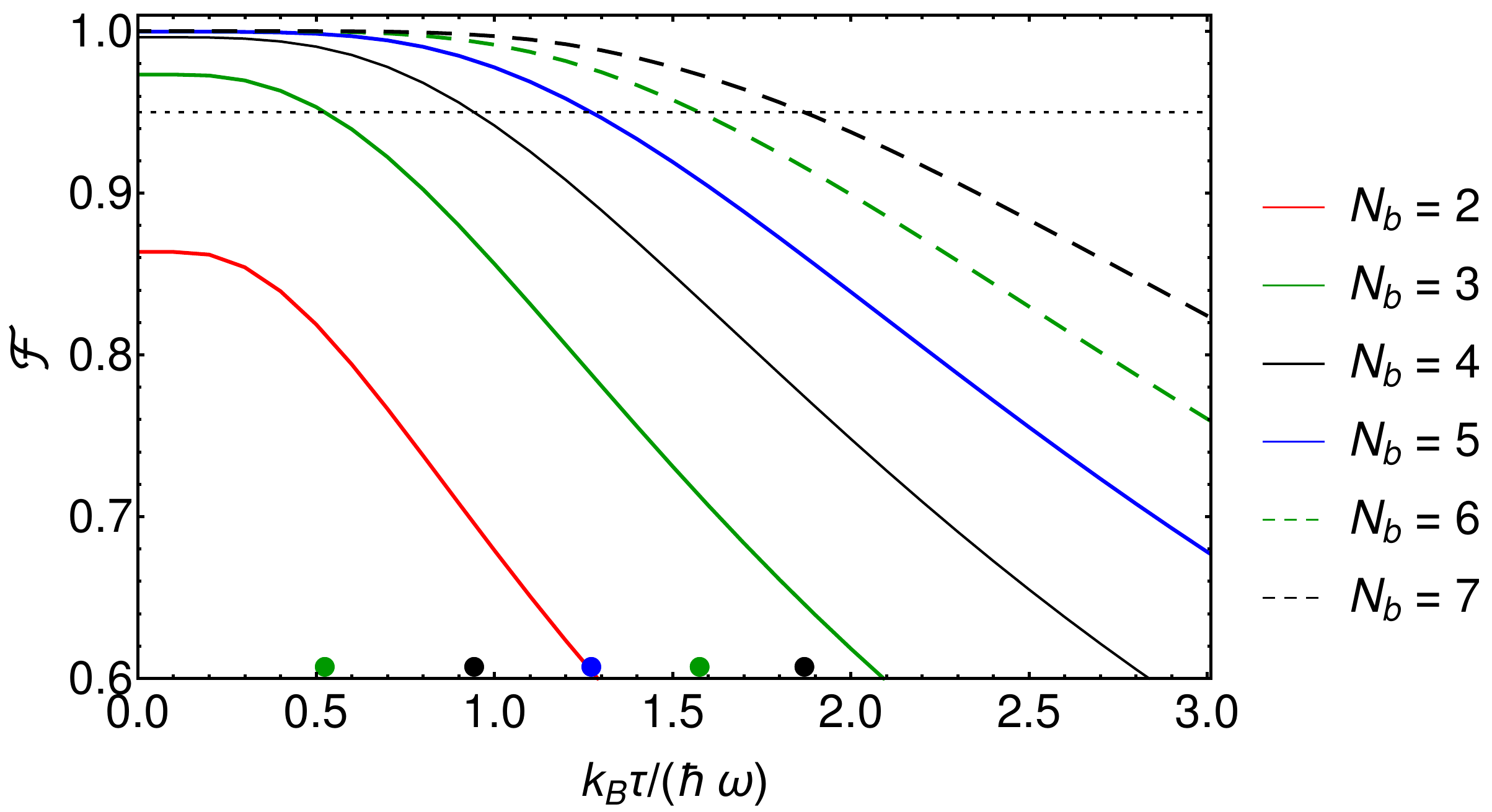}
\end{center}
\caption{
Splitting of a trap from height $h_{i} = 0$ to $h_{f} = 20 \hbar\omega$, sinusoidal scheme, temperature $\tau = 0$.
(a) Fidelity $\Fid$ versus process time $T$ for different $N_b$, $N_p=2$.
(b) Fidelity $\Fid$ versus $N_b$ for different $N_p$, $T =2/\omega$.
(c) Fidelity $\Fid$ versus temperature $\tau$ for different $N_b$, $T = 2/\omega$, $N_p=2$.
In all figures, the horizontal black dotted line indicates a fidelity of $\Fid=0.95$.
\label{Fig6:Split}}
\end{figure*}

\subsection{Splitting}

In our final example we will discuss the process where raising a Gaussian barrier at the center of a harmonic trap leads to a splitting of the atomic cloud. For this we chose  
\begin{equation}
V(x,t) = \frac{1}{2} m \omega^{2}x^{2} + h(t)e^{-x^{2}/d^2} ,
\end{equation}
where again $d=\sqrt{{\hbar}/{m\omega}}$.
The time dependence of the barrier height chosen as
\begin{equation}
h(t) = h_{i} + (h_{f} - h_{i})\sin^{2}\left(\frac{\pi t}{2 T}\right), 
\end{equation}
where $h_i$ is the initial height of the barrier at $x=0$ before the splitting and $h_f$ after the process.
Similarly to the case of expansion, splitting is a symmetric operation , i.e.~the Hamiltonian is commuting with the parity operator. As such it is expected that even numbers of additional particles are more effective than are odd numbers. Splitting is also quite distinct from the other manipulations in that it affects higher energy states in the trap less, whereas transport or expansion affect the whole spectrum of states in the trap.
In the following, we set $h_i = 0$  and $h_f = 20 \hbar\omega$, which lead to a final separation in two wells for approximately the 18 lowest energy eigenstates.

In \fref{Fig6:Split}(a)
one can see that, as expected, increasing $N_b$ gives higher fidelities $\Fid$ on shorter timescales
and $\Fid$ increases monotonically with $T$. In fact, the process is very robust and already for $N_b = 3$ a fidelity of $\Fid \ge 0.95$ can be achieved for almost instant timescales.

The dependence of the fidelity on $N_b$ is shown for different $N_p$ in \fref{Fig6:Split}(b). For odd numbers of particle $N_p$ one can see an effect similar to the one observed in the expansion process, where an even number of buffer particles $N_b$ is needed to see an increase in fidelity. This can again be understood by considering the symmetric nature of the splitting dynamics. However, while one would naively expect the same for states with even numbers of particles $N_p$, it is absent in this case. The reason for this can be found in the specific structure of the eigenspectrum of the split trap, where for our parameters successive even and odd eigenstates are effectively energetically degenerate. An even number of particles in the system therefore has two particles with energies close to the Fermi edge and adding any number of buffer particles will lead to an increase in fidelity as one possible transition is blocked.

Finally, from  \fref{Fig6:Split}(c), one can see that the splitting is slightly more sensitive to temperature than the previous two operations.
The dots on the horizontal axis show heuristically that an additional buffer particle is required for every increase in temperature of about $0.25 k_B/\hbar\omega$, while in the previous two schemes this was about $k_B/\hbar\omega$.

\section{Conclusion}
\label{sec:conclusions}

In this work we have explored the idea of using Pauli blocking for speeding up adiabatic evolution
by  using an additional layer of buffer particles to protect the lowest-energy fermions 
when the system parameters are dynamically changed.
We have presented a thorough investigation, both analytical and numerical, 
showing that the presence of this additional layer allows the speed-up of adiabatic manipulations
without exciting unwanted transitions.
By discussing three different examples, we have demonstrated that this method is robust and applicable to a wide range of scenarios.

The proposed technique is particularly well suited to protect ground states during changes of the external potential, resulting in a speed-up of ground state preparation in potentials for which these states cannot easily be prepared directly with high fidelity. The method does not require precise knowledge of the shape of the trap or the energy spectrum of the system.
It is also insensitive to the details of how the trap parameters are varied in time and no specific time-dependence
of the parameter functions is necessary, which might be very complex and hard to implement experimentally.
All this makes it a very robust and readily applicable technique.

Finally, the presented study gives a deeper insight into the phenomenon of the orthogonality catastrophe.  We have  shown that the fidelity of a subsystem can be much larger than the one of the full many-body system and in particular, that the particles close to the Fermi edge play a much stronger role in the effect of the many-body state becoming orthogonal.

\section*{Acknowledgments}
We are grateful to David Rea for useful discussion and commenting on the manuscript.
TD acknowledges support by the Irish Research Council (GOIPG/2015/3195).
This work was supported by the Okinawa Institute of Science and Technology Graduate University.

\appendix
\begin{widetext}

\section{Calculation of the fidelity}
\label{sec:app}

We calculate the fidelity of the final state, $\Fid = \braket{\Psi}{\Mop\Psi}$, with the measurement operator defined by \eref{eq:M_op}, where $\ket{\Psi}$ is the state of our $N$-fermion wave function after some unitary time evolution.
We want to calculate $\Fid$ as a function of the single-particle states $\ket{\psi_i}$, cf. \eref{eq:multi_wf}.
Expanding the definitions of $\Mop$ and $\ket{\Psi}$, we get
\begin{align*}
\Fid &= \dfrac{1}{N!N_b!} \sum_{\sigma} \sum_{p,q} \sgn(p) \sgn(q) \hspace{1 mm}
\prod_{i=1}^{N_p} \braket{\psi_{p(i)}}{\phi_{\sigma(i)}} \braket{\phi_{\sigma(i)}}{\psi_{q(i)}} \prod_{j=N_p+1}^{N} \braket{\psi_{p(j)}}{\psi_{q(j)}}\\
&= \dfrac{1}{N! N_b!} \sum_{\sigma} \sum_{p,q} \sgn(p) \sgn(q)
\prod_{i=1}^{N_p} \braket{\psi_{p(\sigma^{-1}(i))}}{\phi_{i}} \braket{\phi_{i}}{\psi_{q(\sigma^{-1}(i))}}
\prod_{j=N_p+1}^{N} \braket{\psi_{p(\sigma^{-1}(j))}}{\psi_{q(\sigma^{-1}(j))}} .
\end{align*}
Since the $\ket{\psi_{i}}$ are orthogonal before manipulation (as eigenstates of the Hamiltonian), they remain orthogonal after the unitary evolution. Let us also define $P = p \circ \sigma^{-1}$ and $Q = q \circ \sigma^{-1}$, so that
\begin{equation*}
\Fid = \dfrac{1}{N_b!} \sum_{P,Q} \sgn(P) \sgn(Q)
\prod_{i=1}^{N_p} \braket{\psi_{P(i)}}{\phi_{i}} \braket{\phi_{i}}{\psi_{Q(i)}} \prod_{j=N_p+1}^{N} \delta_{P(j)Q(j)} .
\end{equation*}
We see that only the permutations that fulfill $P(j) = Q(j)$ for $j = N_{p} + 1 ,\ldots,N$ contribute to the sum.
This allows us to rewrite the contributing permutations as $P = \mu \circ \pi_P$ and $Q = \mu \circ \pi_Q$.
$\mu$ should be a permutation $\mu:\{1,\ldots,N\} \to \{1,\ldots,N\}$
with $\mu(i)=P(i)=Q(i)$ for $i>N_p$ and $\mu(i)<\mu(i+1)$ for $i=1,\ldots,N_p-1$.
$\pi_P = \mu^{-1} \circ P$ and $\pi_Q = \mu^{-1} \circ Q$ are then permutations on $\{1,\ldots,N\}$
such that they permute $\{1,\ldots,N_{p}\}$ but act as the identity on $\{N_{p}+1,\ldots,N\}$.
Note that there is a one-to-one correspondence between $P$ and the pair $\mu,\pi_P$.
Then we get
\begin{equation*}
\Fid = \dfrac{1}{N_b!} \sum_\mu \sum_{\pi_P,\pi_Q} \sgn(\pi_P) \sgn(\pi_Q)
\prod_{i=1}^{N_p} \braket{\psi_{\mu(\pi_P(i))}}{\phi_{i}} \braket{\phi_{i}}{\psi_{\mu(\pi_Q(i))}}.
\end{equation*}
This fidelity is independent of $\mu(N_p+1),...\mu(N)$. Therefore, for each $\mu$
we can define a mapping $U: \{1,\ldots,N_{p}\} \to \{1,\ldots,N\}$ by $U(i)=\mu(i)$ for $i=1..N_{p}$
such that $U(i) < U(i+1)$ for $i=1..N_{p}-1$. Note that each $U$ can also be viewed as a subsets of cardinality $N_p$
of the set $\{1,\ldots,N\}$.
As $N_b!$ different $\mu$ result in the same $U$, this allows us to write the fidelity as
\begin{equation*}
\Fid = \sum_{U} \sum_{\pi_P} \sum_{\pi_Q} \sgn(\pi_P) \sgn(\pi_Q)
\prod_{i=1}^{N_p} \braket{\psi_{U(\pi_P(i))}}{\phi_{i}} \braket{\phi_{i}}{\psi_{U(\pi_Q(i))}} 
 =  \sum_{U} \left| \sum_{\pi_P} \sgn(\pi_P) \prod_{i=1}^{N_p} \braket{\psi_{U(\pi_P(i))}}{\phi_{i}}\right|^2
,
\end{equation*}
which corresponds to Eq. (\ref{eq:fid}).
\end{widetext}

\end{document}